\documentclass[12pt]{article}

\catcode`\@=11
\@addtoreset{equation}{section}

\global\arraycolsep=2pt
\usepackage{mathrsfs,amsbsy,amssymb,latexsym,amsfonts,amsmath,cite}
\usepackage{multirow,bigdelim,bm,wrapfig,ascmac}
\usepackage[dvipdfmx]{graphicx}
\usepackage[dvipdfmx]{color}
\allowdisplaybreaks

\newcommand{\dd}{{\rm d}}

\newcommand{\gym}{g_{\text{YM}}}
\newcommand{\gs}{g_{\text{s}}}

\newcommand{\im}{\text{Im}\, }
\newcommand{\pd}{{\cal D} }
\newcommand{\ads}{\text{AdS} }
\newcommand{\tf}{T_{\text{F}} }

\newcommand{\diag}{{\text{diag}} }
\newcommand{\tr}{\text{tr} }

\newcommand{\ex}{\text{ex}}
\newcommand{\eff}{\text{eff}}

\newcommand{\cl}{\text{cl} }

\newcommand{\2}{{\it 2} }

\newcommand{\start}{{(\text{s})} }

\usepackage[dvipdfm]{hyperref}

\begin{document}

\begin{flushright}
\parbox{4.2cm}
{KUNS-2438}
\end{flushright}

\vspace*{2cm}

\begin{center}
{\Large \bf Holographic description of the Schwinger effect \\ 
in electric and magnetic fields}
\vspace*{2cm}\\
{\large Yoshiki Sato\footnote{E-mail:~yoshiki@gauge.scphys.kyoto-u.ac.jp} 
and 
Kentaroh Yoshida\footnote{E-mail:~kyoshida@gauge.scphys.kyoto-u.ac.jp} 
}
\end{center}

\vspace*{1cm}
\begin{center}
{\it Department of Physics, Kyoto University \\ 
Kyoto 606-8502, Japan} 
\end{center}

\vspace{1cm}

\begin{abstract}
We consider a generalization of the holographic Schwinger effect proposed by Semenoff and Zarembo 
to the case with constant electric and magnetic fields. There are two ways to turn on magnetic fields, 
i) the probe D3-brane picture and ii) the string world-sheet picture. 
In the former picture, magnetic fields both perpendicular and parallel to the electric field are activated 
by a Lorentz transformation and a spatial rotation. 
In the latter one, the classical solutions of the string world-sheet corresponding to circular Wilson loops 
are generalized to contain two additional parameters encoding the presence of magnetic fields. 
\end{abstract}

\thispagestyle{empty}
\setcounter{page}{0}

\newpage

\section{Introduction}

In the vacuum of quantum electrodynamics (QED), virtual $e^-$ and $e^+$ pairs are incessantly created and annihilated. 
The pairs can become real particles in the presence of strong electric-field. This realization is well known 
as a novel non-perturbative phenomenon called the Schwinger effect \cite{HE,Schwinger}. 
The production rate $\Gamma $ of the pairs in a homogeneous electric field 
is computed in the weak-coupling and weak-field approximation like 
\[
\Gamma = \frac{(eE)^2}{(2\pi)^3} {\rm e}^{-\frac{\pi m^2}{e E}}\,. 
\] 
It has not been observed in nature yet, but it is expected to be done in the near future \cite{Dunne08,Ringwald} 
(For a recent review on the Schwinger effect, see \cite{Dunne12}).  

\medskip 

The derivation of the pair-production rate was refined to an arbitrary coupling \cite{AAM}. 
This generalization leads to examining the expectation value of a circular Wilson loop in the computational process. 
It can be evaluated using the weak-field condition for an arbitrary coupling and eventually 
gives a small correction. The result is given by \cite{AAM} 
\begin{eqnarray}
\Gamma = \frac{(eE)^2}{(2\pi)^3} {\rm e}^{-\frac{\pi m^2}{e E} + \frac{1}{4}e^2}\,. \label{AAM}
\end{eqnarray}
The similar formula holds for magnetic monopoles \cite{AM}.
Then it is fair to ask what happens when the computation of \eqref{AAM} is applied to 
the AdS/CFT correspondence \cite{M,GKP,W}. 

\medskip 

Some observables in strongly-coupled gauge-theories are computed in terms of 
the classical gravitational-theory via the AdS/CFT correspondence \cite{M,GKP,W}. 
An example is the expectation value of a Wilson loop, which
can be evaluated as the minimal surface of the string world-sheet attaching the boundary \cite{Wilson1,Wilson2}. 
Hence this is an advantage in the holographic setup.  
On the other hand, $U(1)$ gauge theories like QED cannot be argued directly because 
$SU(N)$ ones simply correspond to gravitational theories. Thus one has to consider  
an intricate setup to discuss the Schwinger effect in the holographic scenario. 

\medskip 

The setup we are concerned with is the correspondence between type IIB string theory 
on $\ads _5 \times S^5$ and ${\cal N}=4$ super Yang-Mills (SYM) theory in four dimensions. 
One way to realize a $U(1)$ gauge theory in this setup is to spontaneously break the gauge group 
from $SU(N+1)$ to $SU(N) \times U(1)$\,. Then the computation in the gauge-theory side is almost 
the same as in scalar QED in four dimensions. It is of paramount importance that 
the holographic description is available to evaluate the expectation value of a circular Wilson loop. 
The Schwinger effect in this direction is studied in \cite{GSS,SZ,Ambjorn,BKR}\footnote{For the pair creation of 
open strings in flat space, see \cite{max1,max2}.}. 

\medskip 

In this note we consider a generalization of the scenario by Semenoff and Zarembo \cite{SZ} 
to the pair productions in the presence of electric and magnetic fields. 
There are two ways to turn on magnetic fields, 
i) the probe D3-brane picture and ii) the string world-sheet picture. 
In the former picture, magnetic fields both perpendicular and parallel to the electric field are activated 
by a Lorentz transformation and a spatial rotation. 
In the latter one, the classical solutions of the string world-sheet corresponding to  circular Wilson loops 
are generalized to contain two additional parameters encoding the presence of magnetic fields. 

\medskip 

This note is organized as follows.  
In section 2 we prepare the setup to consider the holographic Schwinger effect. 
In section 3 we consider a generalization of the holographic Schwinger effect 
to the case with electric and magnetic fields. We show two ways to turn on magnetic fields. 
Section 4 is devoted to conclusion and discussion. 

\section{Holographic Schwinger effect}

Let us consider a stack of $N+1$ D3-branes on which the ${\mathcal{N}}=4$ $SU(N+1)$ SYM theory
in four dimensions is realized as the low-energy effective theory. Then separating a single D3-brane 
leads to the Higgs mechanism in the theory.  

\medskip 

The $SU(N+1)$ gauge field $\hat{A}_{\mu}~(\mu=0,\ldots,3)$\,,  
the real scalar fields $\hat{\Phi}_I~(I=1,\ldots,6)$
and the fermion fields $\hat{\Psi}$ are decomposed as follows: 
\begin{equation}
 \hat{A}_{\mu}=
\begin{pmatrix}
A_\mu &\omega _\mu \\
\omega _\mu ^\dagger &a_\mu
\end{pmatrix}\, , \quad 
 \hat{\Phi}_{I}=
\begin{pmatrix}
\Phi_I &\omega _I \\
\omega _I ^\dagger &m\theta _{I}+\phi _I
\end{pmatrix}\, ,\quad
 \hat{\Psi}=
\begin{pmatrix}
\Psi &\chi \\
\chi ^\dagger &\psi
\end{pmatrix}\, . 
\label{decomp}
\end{equation}
Here the diagonal components $A_\mu~[a_{\mu}]$, $\Phi _I~[\phi_I]$, and $\Psi ~[\psi]$ 
are the $SU(N)~[U(1)]$ gauge, the scalar and the fermion fields, respectively. 
The non-diagonal components $\omega _\mu$, $\omega _I$, and $\chi$ transform as the fundamental representation of $SU(N)$ 
and form the W-boson multiplet.
Finally $m\theta _I$  are the vacuum expectation values, where $\theta _I$ satisfy  $\sum_I (\theta _I)^2=1$\,. 
Then $\phi _I$ are regarded as  fluctuations around them.

\medskip 

According to the decomposition 
\eqref{decomp}, 
the original action $S_{{\cal N}=4}^{SU(N+1)}$ is decomposed into the three parts, 
\[
S_{{\cal N}=4}^{SU(N+1)} \quad \longrightarrow \quad S_{{\cal N}=4}^{SU(N)} + S_{{\cal N}=4}^{U(1)} + S_{{\rm W}} \,,
\]
and each of them is described below, 
\begin{eqnarray}
&& 1) \mbox{~~${\mathcal{N}}=4$ $SU(N)$ SYM theory, $S_{{\cal N}=4}^{SU(N)}$}\,, \qquad 
2) \mbox{~~${\mathcal{N}}=4$ $U(1)$ SYM theory, $S_{{\cal N}=4}^{U(1)}$}\,, \nonumber \\ 
&& 3) \mbox{~~the action of the W-boson multiplet,  $S_{\rm W}$}\,. \notag
\end{eqnarray}
For later purpose, we shall write down the concrete expression of $S_{\rm W}$ only, 
\begin{eqnarray}
S_{\text{W}}=\frac{1}{\gym ^2}\int\! \dd ^4x \, \left[(D_\mu \omega _I)^\dagger D^\mu \omega _I
+\omega _I^\dagger (\Phi _K   -m\theta _K)^2\omega _I  
-m^2\omega _I^\dagger \theta _I \theta _J\omega _J +\cdots \right]\,.
\end{eqnarray}
The covariant derivative is defined as  
\begin{eqnarray}
D_\mu \equiv \partial _\mu -iA_\mu +ia_\mu\,, 
\end{eqnarray}
and the ellipsis represents the kinematic terms of the other components of the W-boson multiplet 
and the other interaction terms.

\medskip 

After taking the near-horizon limit, 
a stack of $N$ D3-branes is replaced by the $\ads _5\times S^5$ geometry, 
\begin{eqnarray}
\dd s^2 = \frac{r^2}{L^2} \eta _{\mu \nu }\dd x^\mu \dd x^\nu + \frac{L^2}{r^2}\dd r^2 + L^2 \dd \Omega_{(5)}^2\,, 
\end{eqnarray}
where $\dd \Omega_{(5)}^2$ is the metric of the five-dimensional sphere with unit radius 
and $L$ is the common radius of AdS$_5$ and $S^5$\,. The metric $\eta_{\mu\nu}$ describes the Minkowski spacetime. 
The isolated D3-brane is described as a probe D3-brane extending in AdS$_5$\,. 

\medskip 

In the scenario of the holographic Schwinger effect by Semenoff and Zarembo \cite{SZ}, 
the $U(1)$ gauge field is treated as a constant external field, 
while the $SU(N)$ gauge field is regarded as a dynamical field. 
Then, following the work of  Affleck, Alvarez and Manton \cite{AAM}, 
the production rate  can be evaluated. 
The resulting expression includes the expectation value of a circular Wilson loop of $SU(N)$ gauge field. 
It is a point that one has to evaluate the non-abelian Wilson loop rather than the abelian one 
in comparison to the QED case. Thus it is not an easy task any more in the field-theory framework, 
but now the holographic computation based on the string world-sheet is applicable. 

\subsection{Properties of an electric field from D3-brane action}

Let us first  anticipate the expected behavior of an electric field 
from the point of view of the Dirac-Born-Infeld (DBI) action describing the probe D3-brane. 

\medskip 

The square-root part of the probe D3-brane action is given by 
\[
S _{\text{DBI}}=-T_{\text{D3}} \int\!\dd ^4 x \, \sqrt{-\det (g_{\mu \nu}+{\cal F}_{\mu \nu})}\,, 
\]
where the D3-brane tension $T_{\rm D3}$ and the world-volume flux ${\cal F}_{\mu\nu}$ are, respectively, 
\[
T_{\text{D3}}=\frac{1}{\gs (2\pi )^3 \alpha '^2}\,, \qquad 
{\cal F}_{\mu \nu}\equiv B_{\mu \nu}+2\pi \alpha 'F_{\mu \nu}\,. 
\]
The string tension is given by $T_{\rm F} =1/2\pi\alpha'$ and the string coupling constant is $\gs$\,. 
Then $B_{\mu \nu}$ is the pull-back of the NS-NS two-form $B_\2$
and $F_{\mu \nu}$ is the world-volume flux living on the D3-brane.

\medskip 

\begin{figure}[tbp]
\begin{center}
\includegraphics[scale=.4]{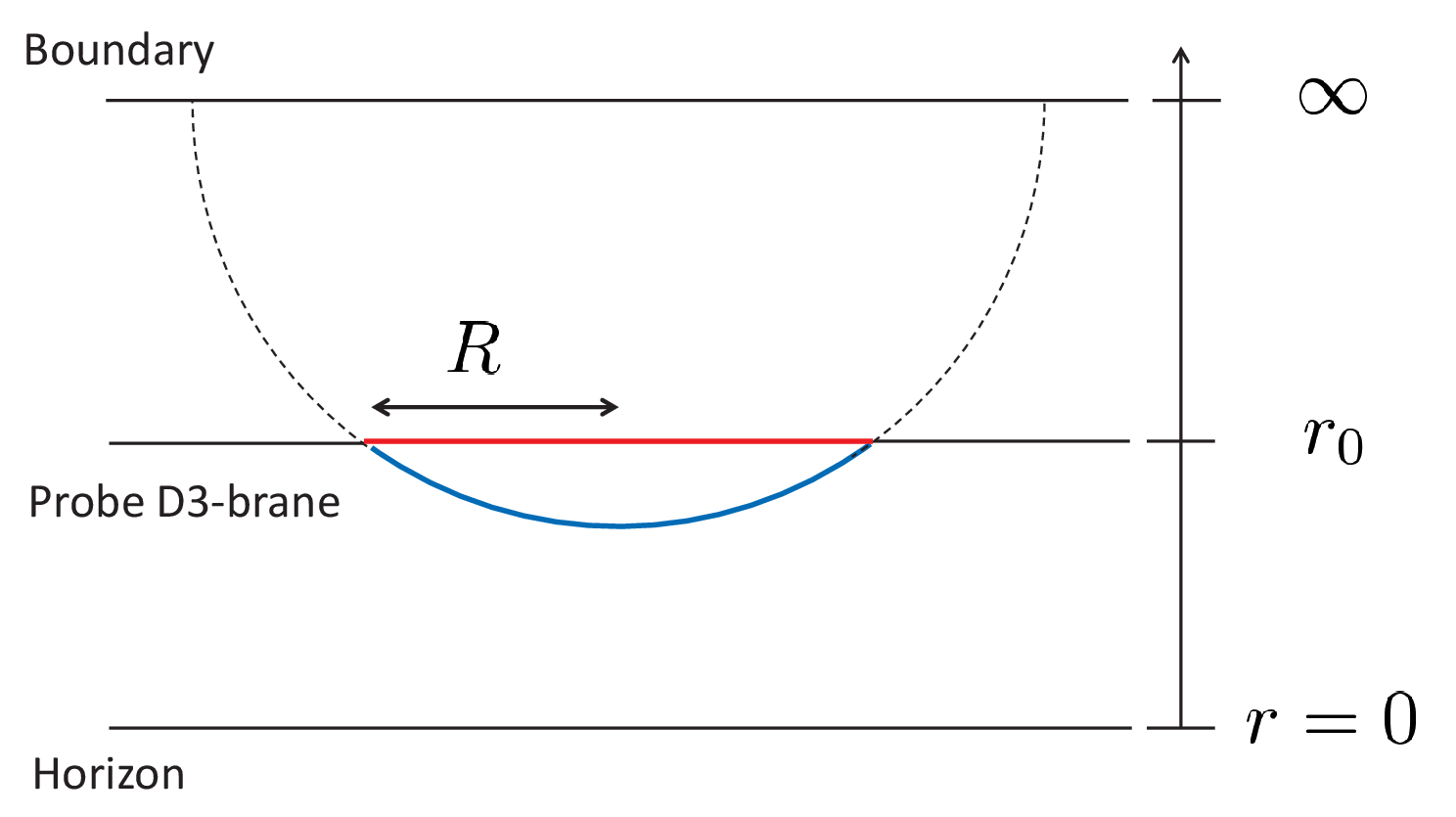}
\end{center}
\vspace*{-0.5cm}
\caption{\footnotesize The location of the probe D3-brane and the configuration of the string world-sheet. \label{configuration:fig}}
\end{figure}

Henceforth we will consider the case that the probe D3-brane is located at $r=r_0$ in the AdS space 
rather than near the boundary, as depicted in Fig.\,\ref{configuration:fig}.  
In addition, an electric field $E$  $(={\cal F}_{01}/2\pi \alpha ')$ is equipped with the probe D3-brane. 
Then the DBI action can be rewritten as  
\begin{align}
S _{\text{DBI}}=-T_{\text{D3}}\frac{r_0^4}{L^4 }\int \!\dd ^4 x\, \sqrt{1-\frac{(2\pi \alpha ')^2L^4}{r_0^4}E^2}\,. 
\label{d3}
\end{align}
When the electric field is given by 
\begin{equation}
E_{\text{DBI}}=\frac{1}{2\pi \alpha '}\frac{r_0^2}{L^2}\,,
\label{ecrit}
\end{equation}
the DBI action vanishes. When the electric field grows more than \eqref{ecrit}, 
the DBI action becomes ill-defined.

\medskip 

Let us write \eqref{ecrit} with the gauge-theory parameters. 
$r_0$ is related with the  mass of the W-boson multiplet, $m$.
The mass  is the energy of a single  string stretching 
between the probe D3-brane at $r=r_0$ and the horizon at $r=0$\,.
The induced metric for the string is $g_{ab}=\diag (-r^2/L^2, L^2/r^2)$ 
and hence the mass is given by 
\begin{equation}
m=\tf \int _{0}^{r_0}\!\dd r\, \sqrt{-\det g_{ab}}\, =\frac{r_0}{2\pi \alpha '}\,.
\label{wmass}
\end{equation}
Using \eqref{wmass} and eliminating $r_0$\,, the electric field \eqref{ecrit} is rewritten as  
\begin{equation}
E_{\text{DBI}}=\frac{2\pi m^2}{\sqrt{\lambda}}\,.
\label{ec}
\end{equation}

\subsection{Pair creation in ${\cal N}=4$ SYM}

It is a turn to focus upon a pair creation of $\omega _I$\,, for simplicity, in the Higgsed 
${\cal N}=4$ $SU(N+1)$ SYM theory in the large $N$ limit. We will work in the Euclidean siganture below, 
unless otherwise stated.  

\medskip 

First of all, let us point out the difference from the QED case \cite{Schwinger,AAM}. 
In the $\mathcal{N}=4$ SYM theory, the number of $\omega _I$ is increased and hence 
the production rate is proportional to $N$\,. Then the loop corrections of the W-boson multiplet and a $U(1)$ photon 
are suppressed in the large $N$ limit because the ones are proportional to $1/N$ and $\gym ^2=\lambda /N$\,, 
respectively. 
The $SU(N)$ gauge field is regarded as a fluctuation and the $U(1)$ gauge field is done
as an external field. That is, the covariant derivative is given by 
\begin{equation}
D_\mu =\partial _\mu -iA_\mu +ia_\mu ^\ex \, .
\end{equation}

The pair-production rate (per unit volume) $\Gamma$ is generally written as 
\[
\Gamma =2 \, \im \varepsilon _0\,,
\]
in terms of the vacuum energy density $\varepsilon _0$\,.
In the ${\cal N}=4$ SYM theory, it is represented by 
\begin{align}
V_4\Gamma &=-2\, \im \ln \int\! \pd A \pd \Phi  \pd \omega \,
{\rm e}^{-S_{{\cal N}=4}^{SU(N)}-S_{\text{W}}} \nonumber \\
 &=-2\, \im \ln \int\! \pd A \pd \Phi \, {\rm e}^{-S_{{\cal N}=4}^{SU(N)}-S_{\eff}} \,,
\end{align}
where the effective action is given by  
\begin{align}
S_{\eff}&=-\ln \int\! \pd \omega \,{\rm e}^{-S_\text{W}} \notag \\
&=\frac{5N}{2}\tr _{\text{op}}\, \tr _{SU(N)}\, 
\ln \left[-D_\mu D^\mu + (\Phi _K   -m\theta _K)^2  \right]\,.
\end{align}
In the last expression, $\tr _{\text{op}}$ is the trace about eigenvalues of  operators and
$\tr _{SU(N)}$ is the trace about $SU(N)$.
The factor $5N/2$ comes from the number of $\omega _I$\,.  

\medskip 

Then the production rate is evaluated as 
\begin{align}
V_4\Gamma &=-2\, \im \ln \int \pd A \pd \Phi \,
{\rm e}^{-S_{{\cal N}=4}^{SU(N)}-S_{\text{eff}}} 
\simeq 2\, \im \langle S_{\eff} \rangle \notag \\
&=5N\, \im \left \langle \tr _{\text{op}}\, \tr _{SU(N)} \, 
\ln \left[-D_\mu D^\mu + \left(\Phi _K   -m\theta _K\right)^2 \right] \right \rangle \, , 
\label{b10}
\end{align}
where the expectation value has been utilized, 
\[
\left\langle g[A, \Phi] \right\rangle =
 \frac{\displaystyle  \int\! \pd A \pd \Phi\,g[A,\Phi ]\, {\rm e}^{ -S_{{\cal N}=4}^{SU(N)}[A,\Phi] } }
{ \displaystyle \int\! \pd A \pd \Phi\, {\rm e}^{-S_{{\cal N}=4}^{SU(N)}[A,\Phi]}}\,.
\]
By using the Schwinger parametrization and the quantum-mechanical path-integral,
the production rate is expressed as 
\begin{align}
V_4\Gamma 
&=-5N\, \im \int_0^\infty \frac{\dd T}{T}\left \langle \tr _{SU(N)}  {\cal P}\exp \left(-\frac{1}{2}D_\mu D^\mu T+ \frac{1}{2}(\Phi _K   -m\theta _K)^2T \right) \right \rangle \nonumber \\
&=-5N\, \im \int_0^\infty \frac{\dd T}{T}\int\! \pd x \, \left \langle \tr _{SU(N)}   {\cal P} 
{\rm e}^{ -S_{\rm inst}} \right\rangle \,,  \notag \\
& S_{\rm inst} \equiv \int _0^T \dd \tau \, \left[ \frac{1}{2}\dot{x}^2 - iA_\mu \dot{x}^\mu +ia_\mu ^\ex \dot{x}^\mu+\frac{1}{2} \left(\Phi _K -m\theta _K \right)^2 \right]\,.
\label{b13}
\end{align}
The above expression contains the path ordering $\cal P$ 
because the $SU(N)$ gauge field is concerned in the present case. 

\medskip 

Then let us evaluate the $T$-integral.
The argument below is the same as the derivation in Appendix A of \cite{DGO}, up to minor modifications.
The integral to be evaluated is 
\begin{equation}
\int _0^\infty \frac{\dd T}{T}\,{\rm e}^{-S_{\rm inst}}\,.
\label{b14}
\end{equation}
The first and fourth terms in (\ref{b13})  are not invariant under the reparametrization of $\tau$\,.

\medskip 

Consider the transformation $\tau \rightarrow \tilde{\tau} (\tau ) $ such that
$\tilde{\tau} (0 )=0 ,\tilde{\tau} (T )=T$ are satisfied.
By defining $c(\tau ) \equiv \dd \tilde{\tau}/\dd \tau $\,, 
the $T$-integral \eqref{b14} is rewritten as the path integral about $c(\tau)$,
\begin{equation}
\int\! \pd c\, \frac{1}{c}\, \exp \left( -\int _0^1 \dd \tau \left[ \frac{1}{2c}\dot{x}^2 
+\frac{c}{2} \left(\Phi _K -m\theta _K \right)^2 \right]\right) \, ,
\label{b15}
\end{equation}
where we have ignored the terms invariant under the reparametrization of $\tau$\,.
The stationary point about $c(\tau)$ is estimated as 
\begin{equation}
c(\tau )\simeq \frac{\sqrt{\dot{x}^2(\tau )}}{m} \,,
\label{b16}
\end{equation}
by assuming that $m$ is very large. 

\medskip 

By the saddle-point approximation about $c(\tau)$, we obtain
\begin{align}
V_4\Gamma &=-5N \, \im \int\! \pd x \, g[x(\tau)]\, {\rm e}^{-S_{\rm particle}[x]}
\left\langle W[x] \right\rangle\,, 
\label{2.18} \\ 
& S_{\rm particle}[x] \equiv m\sqrt{\int_0^1\dd \tau \, \dot{x}^2}  
-i\int_0^1 \dd \tau \, a_\mu ^\ex \dot{x}^\mu\,,  
\\
& W[x]  =  \tr _{SU(N)}{\cal P}
\exp \left(\int\!\dd\tau\left[ i A_\mu (x)\dot{x}^\mu + \Phi_I \theta _I\sqrt{\dot{x}^2} \right]\right)
\label{wil} \,,
\end{align}
where $g[x(\tau)]$ is a functional.
$g[x(\tau)]$ does not influence the classical solution about the steepest descent about $x(\tau)$.
Then the exponential factor in $\Gamma$ is not changed by $g[x(\tau)]$ 
and $W[x]$ is the $SU(N)$ Wilson loop.

\medskip 

Now let us consider how to evaluate \eqref{2.18}. In the QED case \cite{AAM}, 
the exponential part is evaluated with the steepest descent 
by using the instanton solution, and then the expectation value of the Wilson loop is exactly computed because the gauge field 
is $U(1)$\,. In the present case the exponential factor can be evaluated in the same way. The Wilson loop is now non-abelian 
but it is possible to apply the holographic computation. Naively, one may expect that this computation works well.  

 \medskip 

The classical solutions are obtained by solving the equations of motion obtained from $S_{\rm particle}$\,. 
Then they describe the circular motion with the radius $m/E$\,. By putting them into (\ref{wil}), 
its shape is fixed to be circular.  
Since the expectation value of the circular Wilson loop is evaluated as \cite{DGO,BCFM}
\[
\langle W[x] \rangle={\rm e}^{n\sqrt{\lambda}} \qquad (\mbox{wrapped $n$ times}) 
\]
in the holographic computation,  the resulting classical action is given by 
\begin{equation}
S_\cl ^{(n)}=\left( \frac{\pi m^2}{E}-\sqrt{\lambda} \right)n\,. 
\end{equation}
Then the production rate is evaluated as 
\begin{equation}
\Gamma \sim {\rm e}^{-S_\cl ^{(1)}}=\exp \left( -\frac{\pi m^2}{E}+\sqrt{\lambda} \right)\,.
\label{517}
\end{equation}
From this expression, one can read off the critical flux as 
\begin{equation}
E=\frac{\pi m^2}{\sqrt{\lambda}}\,,
\label{5.17}
\end{equation}
so that the production rate is not exponentially suppressed no longer.
However,  the critical value \eqref{5.17} is different from the critical value \eqref{ec} 
estimated from the DBI argument.

\medskip

In the above calculation of the circular Wilson loop, the probe D3-brane is located 
at $r=\infty$ in the radial direction.
This conflicts with the assumption to place the probe D3-brane at $r=r_0$ rather than $r=\infty$\,.
Hereupon Semenoff and Zarembo \cite{SZ} proposed that the exponential factor and the expectation value of the Wilson loop 
should be replaced by the Nambu-Goto (NG) action of a single string attaching on the probe D3-brane 
 at $r=r_0$ and the coupling to  $B_\2$\,. 
That is, the production rate is expected to be 
\begin{gather}
\Gamma \sim {\rm e}^{-S_{\text{st}}}\,,\qquad S_{\text{st}}=S_{\text{NG}}+S_{B_\2} \,, \notag  \\
S_{\text{NG}}\equiv \tf \int\! \dd ^2 \sigma\, \sqrt{\det G_{ab}} \,, \qquad 
S_{B_\2}\equiv -\tf \int\! \dd ^2 \sigma\, B_{\mu \nu} \partial _\tau x^\mu \partial _\sigma x^\nu\,,\notag
\end{gather}
where $G_{ab}$ is the induced metric on the string world-sheet with the coordinates $\sigma^a=(\tau,\sigma)$\,.
This proposal is supported by the coincidence of the critical flux. 

\medskip 

Let us evaluate the production rate by following the Semenoff and Zarembo's proposal. 
We first find the minimal surface of the string with a circular boundary on the probe D3-brane at $r=r_0$\,. 
When the radius of the Wilson loop is $R$ on the probe D3-brane, 
the solutions of the string world-sheet are given by \cite{BCFM} 
\begin{align}
x_0&=x(\sigma) \cos (2n\pi \tau)\,, \qquad x_1=x(\sigma) \sin (2n\pi \tau)\,, \notag \\
 r&=r(\sigma )\,, \qquad x^2+\left(\frac{L^2}{r}\right)^2=R^2+\left(\frac{L^2}{r_0}\right)^2\,.
\label{sol}
\end{align}
A possible representation of the solutions is 
\begin{align}
x(\sigma)=\frac{1}{\cosh \left(2n\pi \sigma \right)}\sqrt{R^2+\left(\frac{L^2}{r_0}\right)^2} \,, \quad
r(\sigma)=\frac{L^2}{\tanh \left( 2n\pi \sigma \right){ \sqrt{R^2+(L^2/r_0)^2}}}  \,.
\end{align}
By putting the solutions \eqref{sol} into the string action, we obtain 
\begin{equation}
S_{\text{NG}} =2n\pi \tf L^2  \left[ \sqrt{\left(\frac{Rr_0}{L^2}\right)^2+1}-1 \right] \,, \qquad
S_{B_\2}=-n\pi ER^2\,,
\end{equation}
where we have defined as 
\[
E \equiv \tf B_{01}\,,  
\] 
and $E$ is interpreted as an electric field
in the gauge-theory side. 

\medskip 

The string boundary condition on the probe D3-brane is the mixed 
(Robin) due to the presence of $B_\2$ like 
\[
{\cal P}_\mu +\tf B_{\mu \nu}\partial _\tau x^\nu \Big| _{r(\tau ,\sigma _0)=r_0}=0 \,, \quad
{\cal P}_\mu \equiv \frac{\delta S_{\text{NG}}}{\delta \left( \partial _\sigma x^\mu \right)}
=\tf \sqrt{\det G} G^{\sigma \sigma} g_{\mu \nu}\partial _\sigma x^\nu \,.
\]
The value of $R$ is determined from this boundary condition,
\begin{equation}
R=\frac{L^2}{r_0}\sqrt{\frac{{\cal E}^2}{E^2}-1}\,, \qquad
{\cal E}=\tf \frac{r_0^2}{L^2}\,.
\label{528}
\end{equation}
Note that the value of $R$ is also determined by taking a variation of the classical action 
with respect to $R$ in \cite{SZ,BKR}. The two ways lead to the identical result.

\medskip 

Then ${\cal E}$ is regarded as the critical value of the electric field so that $R\geq 0$\,.
This is the same as the critical value \eqref{ec} obtained from the argument on the DBI action.
In terms of the gauge-theory parameters, the production rate is given by 
\begin{equation}
\Gamma \sim \exp \left( -\frac{\sqrt{\lambda}}{2}\left[ \sqrt{\frac{{\cal E} } {E}}-\sqrt{\frac{E } {{\cal E}}} \; \right]^2  
\right)\,, \qquad {\cal E}=\frac{2\pi m^2}{\sqrt{\lambda}}\,. 
\end{equation}
The exponential suppression disappears when $E={\cal E}$\,. 
The production rate is approximated to
\begin{equation}
\Gamma \sim \exp \left( -\frac{\pi m^2}{E} +\sqrt{\lambda} \right)\,, 
\label{hol1}
\end{equation}
when $E\ll {\cal E}$\,. This agrees with \eqref{517}.

\section{Turning on magnetic fields}

Let us discuss a generalization of the scenario proposed in \cite{SZ} 
by including magnetic fields both perpendicular and parallel to an electric field. 

\subsection{From the DBI action of the probe D3-brane}

We first consider the case that magnetic fields $B_\parallel$ and $B_\perp$ 
on $x$-direction and $y$-direction, respectively, 
as well as an electric field $E$ on $x$-direction. 
Then the square root in the DBI action of the probe D3-brane is expressed in the Lorentzian signature like  
\begin{align}
&-\det
\begin{pmatrix}
-r_0^2/L^2&-2\pi \alpha 'E & &\\
2\pi \alpha 'E &r_0^2/L^2 & &2\pi \alpha 'B_\perp \\
 &&r_0^2/L^2 &-2\pi \alpha 'B_\parallel\\
 & -2\pi \alpha 'B_\perp&2\pi \alpha 'B_\parallel &r_0^2/L^2
\end{pmatrix} \notag \\
=&\left(\frac{r_0^2}{L^2}\right)^4-(2\pi \alpha ')^2\left(\frac{r_0^2}{L^2}\right)^2
\left(E^2-B_\parallel{}^2-B_\perp {}^2\right)+(2\pi \alpha ')^4E^2B_\parallel{}^2\,.
\end{align}
One can find that the critical value of the electric field exists so that the square root vanishes. 
The critical electric-flux is given by 
\begin{align}
E_{\text{c}}(B_{\perp},B_{\parallel}) & \equiv \frac{r_0^2}{2\pi \alpha ' L^2}\sqrt{1+\frac{B_\perp {}^2}{ (r_0^2/2\pi \alpha ' L^2)^2+B_\parallel{}^2}}
={\cal E}\sqrt{1+\frac{B_\perp {}^2}{ {\cal E}^2+B_\parallel{}^2}}\,, \label{critical}
\end{align}
where ${\cal E}$ is the critical electric-flux without magnetic fields, as we have already seen. 
Note that $E_{\rm c}$ is independent of $B_{\parallel}$ when $B_\perp=0$\,, and grows as $B_\perp$ increases. 

\subsection{How to treat magnetic fields in the holographic description }

\begin{figure}[tbp]
\begin{center}
\includegraphics[scale=.4]{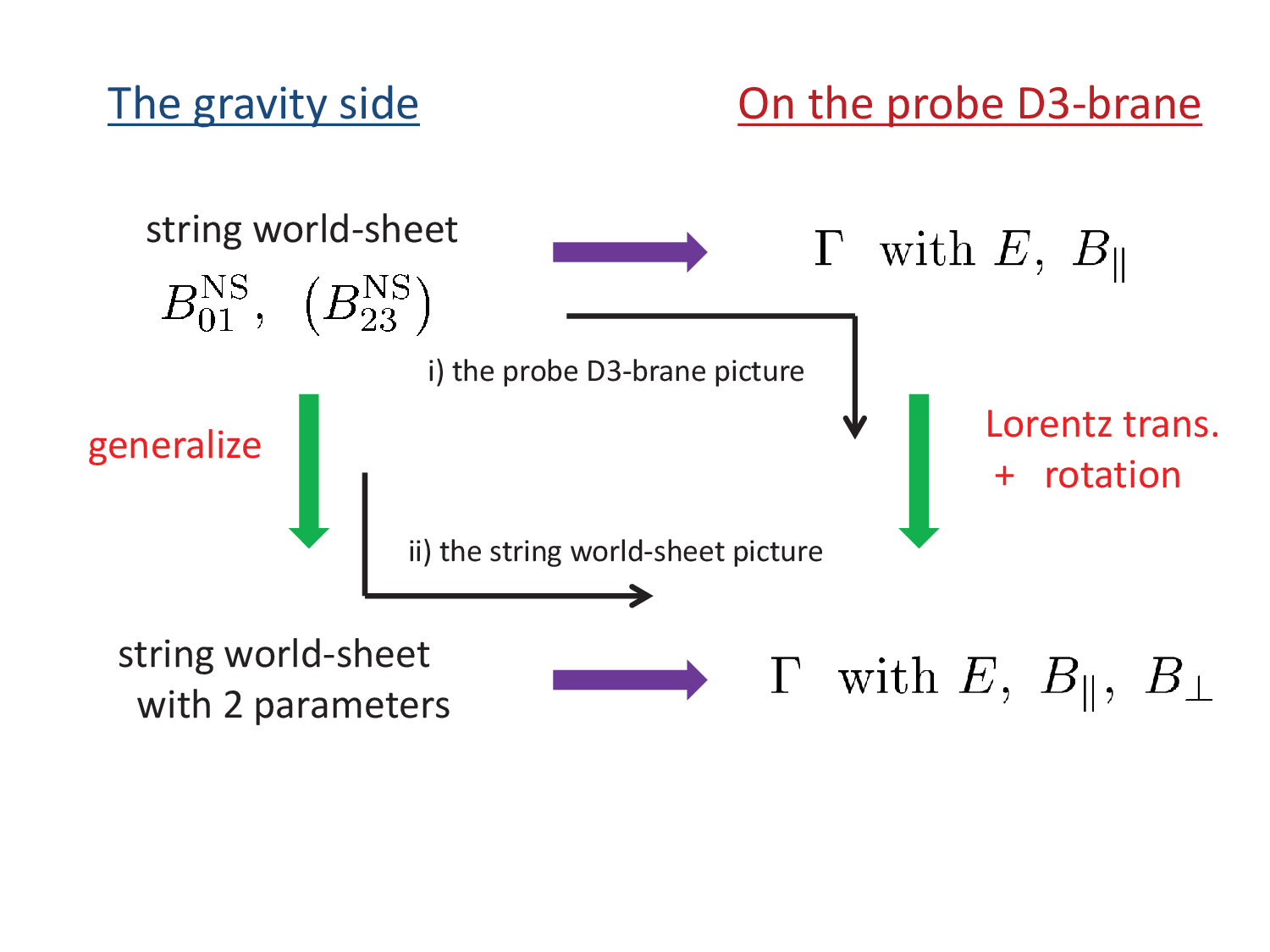}
\end{center}
\vspace*{-2cm}
\caption{\footnotesize The two ways to introduce magnetic fields. \label{outline:fig}}
\end{figure}

The next issue is how one can generalize the holographic description in \cite{SZ} by turning on magnetic fluxes 
so that the critical electric-flux \eqref{critical} is reproduced.  Indeed, there are two ways to treat magnetic fields, 
i) the probe D3-brane picture, ii) the string world-sheet picture, as depicted in Fig.\,\ref{outline:fig}. 

\medskip

In the former picture,  we first consider a circular Wilson loop in 
the presence of  parallel electric and magnetic fields. 
Then
the production rate on the probe D3-brane is computed straightforwardly.  
After that, by performing a Lorentz transformation and a spatial rotation, a perpendicular magnetic field is turned on. 
One can read off the critical electric-flux from the resulting production rate, and it surely agrees with \eqref{critical}. 

\medskip 

In the latter picture, we utilize circular Wilson loop solutions depending on additional parameters,
which are expected to describe magnetic fields.
This approach also leads to the critical flux \eqref{critical}. 

\medskip 

We will explain each of the two ways hereafter. 

\subsubsection*{i) the probe D3-brane picture}

The configuration of the string world-sheet in the presence of parallel electric and magnetic fields  
is the same as the one with the electric field only. 
We first suppose that there are parallel electric and 
magnetic fields on $x$-direction, respectively $E_\start$ and $B_\start$.
By performing the Lorentz boosting in the $z$-direction, $E_\start$ and $B_\start$ are transformed as 
\begin{align}
E'_x=E_\start\gamma , \qquad E'_y=-B_\start\gamma \beta\,, \qquad
B'_x=B_\start \gamma ,\qquad B'_y=E_\start\gamma \beta\,,
\end{align}
where $\beta$ is the Lorentz-boost parameter and 
$\gamma$ is defined as $\gamma \equiv 1/\sqrt{1-\beta ^2}$\,.
The rotation on the $x$-$y$ plane leads to 
\begin{alignat}{3}
E''_x&=E_\start\gamma \cos \vartheta +B_\start\gamma \beta \sin \vartheta , 
\qquad &E''_y&=E_\start\gamma \sin \vartheta -B_\start\gamma \beta \cos \vartheta\,,  \notag \\
B''_x&=-E_\start\gamma \beta \sin \vartheta +B_\start\gamma \cos \vartheta ,
\qquad &B''_y&=E_\start\gamma \beta \cos \vartheta +B_\start\gamma \sin \vartheta\,.
\end{alignat}
Here let us impose 
\[
\tan \vartheta =\frac{\beta B_\start}{E_\start}\,,
\]
 so as to set $E''_y=0$\,, 
and introduce the following notation, 
\[
E=E''_x \, ,\quad B_\parallel =B''_x \, ,\quad B_\perp =B''_y\,.
\]
Solving about $E_\start$, we obtain
\begin{equation}
E_\start^2=\frac{1}{2}\left( E^2-{B_\parallel {}^2}-B_\perp {}^2
+\sqrt{\left(E^2-B_\parallel {}^2-{B_\perp {}^2}\right)^2 +4E^2B_\parallel {}^2}\,\right)\,.
\label{e0}
\end{equation}
Thus, from the computation of the production rate in the presence of the parallel electric and magnetic fields only, 
it is possible to calculate the production rate including the perpendicular magnetic field as well as the parallel electric and 
magnetic fields. 

\medskip 

In the world-sheet description, we know the critical flux in the presence of the electric field only. 
By using this result, the critical value of the electric field in the presence of magnetic fields 
can be derived from the following relation,  
\begin{equation}
{\cal E}^2=\frac{1}{2}\left( E_{\text{c}}^2-{B_\parallel {}^2}-B_\perp {}^2
+\sqrt{\left(E_{\text{c}}^2-B_\parallel {}^2-{B_\perp {}^2}\right)^2 +4E_{\text{c}}^2B_\parallel {}^2}\,\right)\,.
\end{equation}
By solving about $E_{\text{c}}$ again, 
$E_{\text{c}}$ is obtained as a function of ${\cal E}\,, B_\perp$ and $ B_\parallel \,,$
\begin{align}
E_{\text{c}}={\cal E}\sqrt{1+\frac{B_\perp {}^2}{ {\cal E}^2+B_\parallel{}^2}}\,.
\label{3.7}
\end{align}
This expression agrees with the expectation from the argument on the DBI action. 

%%%%%%%%%%%%%%%%%%%%%new
\medskip

As a side note, let us see the allowed range of $B_{\perp}/E$\,. It depends on the presence of $B_{\parallel}$\,. 
The Lorentz-boost parameter $\beta$ is now expressed as a function of $E\,, B_\parallel \,,$ and $B_\perp$\,,
\begin{equation}
\beta ^2
=\frac{E^2+{B_\parallel {}^2}+B_\perp {}^2
-\sqrt{\left(E^2-B_\parallel {}^2-{B_\perp {}^2}\right)^2 +4E^2B_\parallel {}^2}}
{E^2+{B_\parallel {}^2}+B_\perp {}^2
+\sqrt{\left(E^2-B_\parallel {}^2-{B_\perp {}^2}\right)^2 +4E^2B_\parallel {}^2}} <1\,.
\label{0.5}
\end{equation}
When $B_\parallel =0$\,, $\beta =B_\perp /E$ and hence $B_\perp$ is always smaller than $E$\,, 
in particular $B_{\perp} < E_{\rm c}$\,. However, when $B_\parallel \neq 0$\,, 
$B_\perp $ may be greater than $E$\,. One can check it from the relation
%\begin{align}
%E&=\gamma \sqrt{E_\start^2+B_\start^2\beta ^2} \,, \qquad 
%B_\perp =\gamma \beta \frac{E_\start^2+B_\start^2}{\sqrt{E_\start^2+B_\start^2\beta ^2} }\\
%\frac{B_\perp}{E}&=\frac{E_\start^2+B_\start^2}{E_\start^2/\beta +B_\start^2\beta}
%\end{align}
\begin{equation}
\frac{B_\perp}{E}=\frac{E_\start^2+B_\start^2}{E_\start^2/\beta +B_\start^2\beta}\,,
\end{equation}
and the ratio $B_{\perp}/E_{\rm c}$ can be greater than 1 depending on the values of 
$E_\start \,, B_\start \,,$ and $\beta$\,.
%%%%%%%%%%%%%%%%%%%%%%%%%%%%%%%%%%%%%%%%%%%%%%%%%%

\medskip 

By using (\ref{3.7}), the production rate is given by 
\begin{equation}
\Gamma \sim \exp 
\left( -\frac{\sqrt{\lambda}}{2} \left[ 
\sqrt{\frac{{\cal E} } {E_\start}}-\sqrt{\frac{E_\start } {{\cal E}}} \; \right]^2\right)\,,
\end{equation}
where $E_\start$ is given by \eqref{e0}.

\medskip 

Note that this result may be modified by magnetic fields when we take account of the one-loop contributions of $x(\tau)$ 
in the gauge-theory computation. It has not succeeded yet to compute the correct one-loop contributions, 
though some trials have been done \cite{Ambjorn,KM}.

\subsubsection*{ii) the string world-sheet picture}

There is another way to introduce magnetic fields. We shall persist in the gravity side, 
instead of relying on the probe D3-brane description. Naively thinking, 
the magnetic fields in the gauge-theory side are induced by turning on 
$B_\2$ like $B_{23}$ or $B_{31}$ in the gravity side. However, the coupling to $B_\2$ 
vanishes except $B_{01}$ by putting the classical solutions because $x_2=x_3=0$\,. 
Hence one has to seek for an ingenious way to encode the information of magnetic fields 
into the string world-sheet.  

\medskip 

One prescription is to generalize the classical solutions of the string world-sheet 
corresponding to circular Wilson loops \cite{BCFM} so as to 
depend on additional constant parameters. Such an example is given by \cite{MY}
\begin{align}
x_0=-\frac{\ell \sqrt{1-\alpha ^2}\cos \sigma }{\cosh \tau +\alpha \cos \sigma}  \,, \quad
x_1=\frac{\ell \sinh \tau}{\cosh \tau +\alpha \cos \sigma} \,, \quad
r=L^2\frac{\cosh \tau +\alpha \cos \sigma}{\ell \sin \sigma} \,, 
\label{MY}
\end{align}
in the one-patch notation.
Here the constant parameter $\alpha$ satisfies $0\leq \alpha \leq1$\,. 
For $0\leq \alpha < 1$\,, the solutions in \eqref{MY} correspond to circular Wilson loops with the radius 
$\ell/\sqrt{1-\alpha^2}$\,. This is obvious by noting the relation, 
\begin{equation}
x_0^2+ \left(x_1+\frac{\ell \alpha }{\sqrt{1-\alpha ^2}}\right)^2+\left(\frac{L^2}{r}\right)^2
=\frac{\ell ^2}{1-\alpha ^2}\,.
\label{MY2}
\end{equation}
The case with $\alpha =1$\,, in which the radius is divergent, corresponds to the straight line.  
Thus the solutions in \eqref{MY} interpolate between circular Wilson loops 
and the straight Wilson line.

\medskip

Using the solutions in \eqref{MY}, the NG part and the coupling to $B_\2$ in the classical string action 
are evaluated as, respectively,  
\begin{align}
S_{\text{NG}}=2n\pi \tf L^2 \left( \frac{r_0 \ell}{L^2\sqrt{1-\alpha ^2}}-1\right)  \, ,\quad
S_{B_\2}=-n\pi \tf B_{01}\left[ \frac{\ell ^2}{1-\alpha ^2}-\left(\frac{L^2}{r_0}\right)^2\right]\,. 
\end{align}

\medskip 

Then $\ell $ is determined by the boundary condition on the probe D3-brane like  
\begin{eqnarray}
\ell = \frac{r_0}{B_{01}}\sqrt{1-\alpha^2}\,.
\label{l}
\end{eqnarray}
On the other hand, one can read off from \eqref{MY2} that the radius $R$ of the string world-sheet 
on the probe D3-brane is represented by 
\begin{eqnarray}
R^2=\left( \frac{r_0}{B_{01}}\right)^2-\left(\frac{L^2}{r_0}\right)^2\,, 
\label{R}
\end{eqnarray}
where we have used \eqref{l}\,. By using \eqref{R}\,, the classical action is evaluated as 
\begin{align}
S_\cl ^{(n)}
&=n\pi \tf L^2 \left[ \sqrt{\frac{r_0^2}{B_{01}L^2}}-\sqrt{\frac{B_{01}L^2}{r_0^2}} \; \right]^2\,. 
\label{yoso}
\end{align}
Then the critical value is given by $B_{01}^{\rm (cr)}=r_0^2/L^2$ 
so that $S_\cl ^{(n)}=0$ and $R=0$\,.

\medskip

From \eqref{yoso}\,, one can read off the relation between $B_{01}$ and the electric field $E$ 
so as to reproduce the critical flux \eqref{critical} with $B_{\parallel} =0$ as follows:
\begin{eqnarray}
\tf B_{01} =\sqrt{E^2-B_\perp{}^2}\,. 
\end{eqnarray}
This expression implies that $B_{01}$ should be identified with the Lorentz-boosted electric flux. 
According to this identification, the other parameters contained in the solutions \eqref{MY} 
are translated in terms of the gauge-theory language like 
\begin{align}
\ell =\frac{m}{E}\,, \qquad \alpha =\frac{B_\perp}{E} \equiv \beta\,. 
\label{dictionary}
\end{align}
These relations are fixed from the consistency to \eqref{l}\,. 
Thus the solutions \eqref{MY} contain the information of a magnetic field perpendicular to the electric field. 

\medskip 

It is fair to ask how one can include a magnetic field parallel to the electric field into the classical solutions. 
When $B_{\parallel} \neq 0$\,, it is quite natural from the above argument to identify $B_{01}$ as follows: 
\begin{eqnarray}
\tf B_{01} =  \frac{1}{\sqrt{2}}\left[ E^2-{B_\parallel {}^2}-B_\perp {}^2
+\sqrt{\left(E^2-B_\parallel {}^2-{B_\perp {}^2}\right)^2 +4E^2B_\parallel {}^2}\,\right]^{1/2}\,,
\end{eqnarray}
so that the critical flux agrees with the DBI result \eqref{critical}\,. Note that $\tf B_{01} =  E$ consistently 
when $B_{\parallel} \neq 0$ and $B_{\perp} =0$\,. 

\medskip 

According to this identification, the classical solutions in \eqref{MY} should be modified. 
As is expected from the argument in the previous picture, 
the modification is so small that the parameter $1-\alpha^2$ is replaced as 
\begin{eqnarray}
1-\alpha^2 \quad \longrightarrow \quad  \frac{1-\alpha^2 -\delta^2 + \sqrt{(1-\alpha^2 - \delta^2)^2 + 4\delta^2}}{2}\,. 
\end{eqnarray}
Then the boundary is satisfied under the following identification,
\begin{eqnarray}
\delta = \frac{B_{\parallel}}{E}\,. 
\end{eqnarray} 
as well as the previous relations in \eqref{dictionary}\,. 
When $\alpha=0$\,, the parameter $\delta$ is irrelevant to the behavior of the classical solution. 
However, when $\alpha$ takes a finite value, the solution tends to describe the straight Wilson line again 
in the limit $\delta \to \infty$\,.  

\section{Conclusion and discussion}

We have studied the holographic Schwinger effect in the presence of electric and magnetic fields. 
There are two ways to turn on magnetic fields, 
i) the probe D3-brane picture and ii) the string world-sheet picture. 
In the former picture, magnetic fields both perpendicular and parallel to the electric field are activated 
by a Lorentz transformation and a spatial rotation. 
In the latter one, the classical solutions of the string world-sheet corresponding to  circular Wilson loops 
are generalized to contain two additional parameters encoding the presence of magnetic fields. 

\medskip 

In this note we have considered only homogeneous fields. 
It would be an interesting direction to investigate the case with inhomogeneous fields.  
The production rate in the presence of inhomogeneous fields is computed in some $U(1)$ gauge theories \cite{Dunne05}. 
The remaining task is to find out the corresponding solutions of the string world-sheet.  

\medskip 

It would also be very interesting to consider the Schwinger effect for non-abelian gauge fields \cite{na1,na2,na3}
in the holographic QCD frameworks such as the Sakai-Sugimoto models \cite{SS}. 
Our result may find out some applications in this direction. 

\subsection*{Acknowledgments}

We would like to thank T.~Kunihiro, S.~Nakamura and H.~Suganuma  for useful discussions. 
The work of KY was supported by the scientific grants from the Ministry of Education, Culture, Sports, Science 
and Technology (MEXT) of Japan (No.\,22740160). This work was also supported in part by the Grant-in-Aid 
for the Global COE Program ``The Next Generation of Physics, Spun 
from Universality and Emergence'' from MEXT, Japan.

\end{document}